\documentstyle[12pt]{article}
\renewcommand {\c}  {\'{c}}
\newcommand {\cc} {\v{c}}
\newcommand   {\s}  {\v{s}}
\begin{document}
\pagestyle{empty}
\vspace* {13mm}
\baselineskip = 24pt

%************************TITLE PAGE***************************************
%
\begin{center}
{\bf PARTITION FUNCTIONS FOR GENERAL MULTI-LEVEL SYSTEMS}
\\[8.0mm]

S.Meljanac$^{+,1}$ , M.Stoji\c $^{+}$  and D.Svrtan$^{++,2}$ \\[7mm]
$^{+}$ Rudjer Bo\s kovi\c $\,$ Institute, Bijeni\cc ka c.54,\\ 10001 Zagreb,
Croatia\\[3mm] 
$^{++}$ Department of Mathematics, University of Zagreb, Bijeni\cc ka c.30, 
\\ 10001 Zagreb, Croatia \\  
$^{1}$ e-mail: meljanac@thphys.irb.hr\\
$^{2}$ e-mail: dsvrtan@cromath.math.hr
\end{center}

\bigskip
%*********************ABSTRACT******************************************
{\bf Abstract}

We describe a unified approach to calculating the partition functions 
of a general multi-level system with a free Hamiltonian. Particularly,
we present new results for parastatistical systems of any order in
 the second quantized approach. Anyonic-
like systems are briefly discussed.

%**********************************************************************
%
%
%************************MAIN TEXT************************************
%
\newpage
\setcounter{page}{1}
\pagestyle{plain}
\def\leer{\vspace{5mm}}
\baselineskip=24pt
\setcounter{equation}{0}

{\bf Introduction}. Statistics plays a fundamental role in determining macroscopic 
or thermodynamic properties of a quantum many-body system.So far there has been firm 
experimental 
evidence only for two statistics:Bose-Einstein and Fermi-Dirac.Much  effort has been 
made to construct theories with generalized statistics and to search for 
deviations from Bose and Fermi statistics.
Parastatistics [1-3] were the first consistent generalizations of Bose and Fermi statistics 
 in any spacetime dimension.They are  invariant under the permutation group.The N-particle 
states can occur in a large class of representations (REP's) of the permutation group 
$S_{N}$,i.e. not only in totally symmetric REP (for bosons) or totally antisymmetric 
REP (for fermions).Parastatistics can be related to theories obeying ordinary (Bose,Fermi) 
statistics and carrying suitable internal symmetries [3].Hence a natural question arises 
whether there are systematic ways of distinguishing between the two classes of theories.\\
 Other generalized statistics have also been suggested,such as infinite quon 
statistics [4,5],a new version of parastatistics [6],fractional anyonic (braid) statistics [7,8]
 and 
other generalized statistics related to algebras of creation and annihilation operators with 
Fock-like representations [9-12].\\
To investigate physical consequences of such generalized systems,it is important to know 
the corresponding partition functions.There are few partial results for some parastatistical 
systems [13,14].Our motivation is to present a unified approach to calculating the partition 
function of a general multi-level system with a free Hamiltonian.Particularly,we present new results 
for parastatistics and a correction to results for two-level systems for parafermions given in [13]. 
Anyonic-like operator algebras are briefly discussed.

{\bf The partition function for a general multi-level free system}. The statistical average of an 
observable,described by an operator ${\cal {O}}$ in a given ensemble, is defined as
\begin{equation}
 <{\cal O}>=\frac{Tr  {\cal O}\,e^{-\beta H}}{Tr e^{-\beta H}}=\frac{1}{{\cal Z}}
 Tr {\cal O} e^{-\beta H}.
\end{equation}
Here ${\cal Z}$ is the thermodynamic partition function for a multi-level system described by 
M independent creation (annihilation) operators $a^{\dagger}_{i}$ ($a_i$),$i=1,2,..M.$
The operator algebra is defined by a normally ordered expansion $\Gamma$ (generally, 
no symmetries are assumed) [11]:
 \begin{equation}
a_ia^{\dagger}_{j}=\Gamma_{ij}(a^{\dagger},a),
\end{equation}
\noindent
with well-defined number operators $N_i$,i.e. $[N_{i},a_{j}^{\dagger}]=a_{i}^{\dagger}\delta_{ij}$,
$[N_{i},a_{j}]=-a_{i}\delta_{ij}$ and $[N_{i},N_{j}]=0$ for $i,j=1,2,..M$. We assume that there 
is a unique vacuum $|0>$ and the corresponding Fock-like representation.\\
The scalar product (bilinear form) is uniquely defined by $<0|0>=1$ , the vacuum condition 
$ a_i|0> = 0 $, $i=1,2,..,M $ and Eq.(2). A general N-particle state is a linear combination 
of   monomial state vectors ($a^{\dagger}_{i_{1}}\cdots a^{\dagger}_{i_{N}}|0>$),
$i_{1},\cdots ,i_{N}=1,2,...M.$

We consider Fock representations with no state vector of negative squared norm 
but we might have 
norm zero vectors.These null-vectors will represent the only relations  
 between the creation (annihilation) operators on the associated (non degenerate) quotient 
 Fock space .\\
For fixed sequence of indices $i_{1}\cdots i_{N}$, we write its type as $1^{n_{1}}2^{n_{2}}...M^{n_{M}}$,
where $n_{1},n_{2},...,n_{M}$ are multiplicities of appearance $1,2,...M$ in $i_{1} \cdots i_{N}$. 
Note that $n_i \geq 0$ and 
$\sum_{i=1}^{M} n_{i}=N$.Then, there are $N!/{n_{1}!n_{2}!...n_{M}!}$ (in principle) 
different states \\ 
$(a^{\dagger}_{i_{1}}\cdots a^{\dagger}_{i_{N}}|0>)$labelled by all sequences 
of type $1^{n_{1}}2^{n_{2}}...M^{n_{M}}$ and let ${{\cal A}^{\Gamma}}(n_1,\cdots n_M)$ be the
Gram matrix of their scalar products.The number of linearly independent states 
among them is given by $d_{n_{1},\cdots n_{M}}^{ \Gamma}={\it rank}[{{\cal A}^{\Gamma}}(n_1,\cdots n_M)]$,
satisfying
\begin{equation}
0 \,\leq d^{\ \Gamma}_{n_{1},\cdots, n_{M}}\, \leq \frac{N!}{n_{1}!n_{2}!...n_{M}!}.
\end{equation}
All the quantities $d_{n_{1},\cdots n_{M}}^\Gamma$ completely characterize the 
statistics,the partition 
function and the thermodynamic properties of the above free system realized on the corresponding 
Fock space.The free Hamiltonian is then
\begin{equation}
H_0=\sum_{i=1}^{M} E_i \, N_i \, ,
\end{equation}
\noindent
where $E_i$ is the energy of the $i^{th}$ level and $N_i$ are the number operators 
counting particles on the $i^{th}$ level.(Note that the statistics,i.e. the numbers
$d_{n_{1},\cdots n_{M}}$ do not determine uniquely the whole operator algebra,Eq.(2).)
The partition function for a free system described by the $\Gamma$ algebra,
Eq.(2),is given by 
\begin{equation}
{\cal Z}^{\Gamma}(x_1,...x_M)= Tr \,e^{-\beta H_0}= \sum_{N=0}^{\infty} {\cal Z}^{\Gamma}_{N}(x_1,...x_M)\, ,
\end{equation}
\noindent
where
\begin{equation}
{\cal Z}^{\Gamma}_{N}(x_1,...x_M)=\sum_{n_{1}+\cdots +n_{M}=N} d_{n_{1},\cdots n_{M}}^{\Gamma}\,
x_1^{n_{1}}\cdots x_M^{n_{M}}.
\end{equation}
\noindent
Here $d_{n_{1},\cdots n_{M}}^{\Gamma}$ can be considered as the  degeneracy of the state with the energy 
$E=\sum_{i=1}^{M} n_iE_i$ ,where $x_i=e^{-\beta E_{i}}$,$\beta=\frac{1}{kT}$.\\
The central problem is how to compute $d_{n_{1},\cdots n_{M}}^{\ \Gamma}$ , $ n_{1},\cdots, 
n_{M}\geq 0 $.The number of all allowed N-body states distributed over M energy levels is 
given by 
\begin{equation}
D^{\Gamma}(M,N)= \sum_{n_{1}+\cdots +n_{M}=N} d^{\ \Gamma}_{n_{1},\cdots n_{M}}\equiv 
{\cal Z}^{\Gamma}_{N}(\underbrace {1,1,...,1}_{M}).
\end{equation}

{\bf The partition function for permutation invariant multi-level systems}.If the relations  $\Gamma$ in Eq.(2) are invariant under the permutation group 
$S_M$,then the matrix ${\cal A}^{\Gamma}(n_1,\cdots n_M)$ and its rank 
$d^{\ \Gamma}_{n_{1},\cdots n_{M}}$ depend only on the collection of the multiplicities 
$\{ n_{1},\cdots, n_{M} \}$,which written in the descending order $\lambda_1 \geq \lambda_2 \geq 
...\geq \lambda_M \geq 0$,$|\lambda|=\sum_{i=1}^{M}\lambda_{i}=N$, give rise to a 
partition $\lambda$ of $N$,i.e. $d^{\ \Gamma}_{n_{1},\cdots n_{M}}=d^{\ \Gamma}_{\lambda}$   and 
${\cal A}^{\Gamma}(n_1,\cdots n_M)={\cal A}^{\Gamma}_{\lambda}$.\\
If $\lambda_1 = \lambda_2= \cdots  =\lambda_N =1$, $\lambda_{N+1}=\cdots =\lambda_M=0$,
the correspponding Young
diagram,denoted by ${1^N}$, is a column of N boxes. The corresponding $N! \times N!$ 
 ${\it generic}$ matrix is denoted by ${\cal A}^{\Gamma}_{1^N}$.All other matrices ${\cal A}^{\Gamma}_{\lambda}$ ,
 $|\lambda|=N$ for any partition $\lambda $ of N are easily obtained from the generic 
 matrix ${\cal A}^{\Gamma}_{1^N}$.By symmetry of $\Gamma$, 
 it follows that ${\cal A}^{\Gamma}_{1^N}$ can be written as 
\begin{equation} 
{\cal A}^{\Gamma}_{1^N}=\sum_{\pi \in S_{N}} c^{\Gamma}_\pi\, R(\pi) \, ,
\end{equation}
\noindent
where $R$ is the right regular representation of the permutation group $S_N$  and 
$c^{\Gamma}_\pi$ are complex numbers. If the operators $a_i$ and $a^{\dagger}_i$ are adjoint to each other, 
then the matrix ${\cal A}^{\Gamma}_{1^N}$ is hermitian. Of course, 
$ d^{\Gamma}_{1^N}=rank [{\cal A}^{\Gamma}_{1^N}] \leq N!$.
The matrix ${\cal A}^{\Gamma}_{1^N}$ commutes with every permutation matrix 
in the left regular representation. Hence the nondegenerate quotient Fock 
space splits into the sum of irreducible representations (IRREP's) of $S_N$, 
and we can write 
\begin{equation}
d^{\Gamma}_{1^N}=\sum_{\mu} n^{\Gamma}(\mu)\, K_{\mu,1^N} \, ,
\end{equation}
\noindent
where $n^{\Gamma}(\mu)\geq 0$ is multiplicity 
 and $K_{\mu,1^N}$ (Kostka-Foulkes number) is the dimension of the
IRREP $\mu$ of the group $S_N$. The sum in Eq.(9) runs over all partitions $\mu$ 
of N, i.e. all IRREP's. The  multiplicities $n^{\Gamma}(\mu)$ can be 
determined from the spectrum  of the matrix  ${\cal A}^{\Gamma}_{1^N}$.
%Note that ${\cal A}_{\lambda}$ is the same for all $M\geq l(\lambda)$ 
%(l($\lambda $) represents the number of rows in the Young tableaux).
>From Eq.(9) and $S_M$ invariant structure of partition function 
${\cal Z}^{\Gamma}_{N}(x_1,...x_M)$ it follows that
\begin{equation}
d^{\Gamma}_{\lambda}=\sum_{\mu} n^{\Gamma}(\mu)\, K_{\mu,\lambda} \, ,
\end{equation} 
\noindent
with the same $n^{\Gamma}(\mu)$ as in Eq.(9) and where $K_{\mu,\lambda}$ are 
the Kostka-Foulkes numbers enumerating  semistandard(column strict) 
tableaux of weight $\lambda$ and shape $\mu$.Clearly, 
$K_{\mu,\lambda}\leq K_{\mu,1^N} =dim{ \mu} $.The set of all 
$n^{\Gamma}(\mu)$  completely determine the statistics, the partition function and the thermodynamic 
properties of the free $S_M$ invariant system.
The $S_M$ invariant partition function,Eq.(6),becomes 
\begin{eqnarray}
{\cal Z}^{\Gamma}_{N}(x_1,...x_M) & = &\sum_{\lambda;|\lambda|=N} d^{\Gamma}_{\lambda}\,
\sum_{\pi \in S_M} x_1^{\lambda_{1}}\cdots x_M^{\lambda_{M}}= 
\sum_{\lambda;|\lambda|=N} d^{\Gamma}_{\lambda}\, m_{\lambda}(x_1,...x_M) \nonumber \\
&=&\sum_{\mu;|\mu|=N} n^{\Gamma}(\mu) s_{\mu}(x_1,...x_M) ,
\end{eqnarray}
\noindent
where $m_{\lambda}(x_1,...x_M)$ is the monomial $S_M$ symmetric function (summed over all 
distinct permutations of ($\lambda_{1},...\lambda_{M}$),and 
$s_{\mu}(x_1,...x_M)$ are the Schur's functions [15],satisfying  
\begin{equation}
s_{\mu}(x_1,...x_M)=\sum_{\mu}K_{\mu,\lambda}\, m_{\lambda}(x_1,...x_M)
\end{equation} 
 Note that for the numbers $D^{\Gamma}(M,N)$ of all allowable N-particle states
of a permutation invariant M-level system we have:\\
\begin{eqnarray}
D^{\Gamma}(M,N)&= &\sum_{\lambda;|\lambda|=N}d^{\Gamma}_{\lambda} m_{\lambda}(\underbrace {1,1,...1}_{M})
=\sum_{\mu;|\mu|=N} n^{\Gamma}(\mu) s_{\mu}(\underbrace {1,1,...1}_{M})\nonumber \\
& = & \sum_{\mu;|\mu|=N} n^{\Gamma}(\mu) dim \{ \mu \} 
\end{eqnarray}
\noindent
since $s_{\mu}(\underbrace {1,1,...1}_{M})= dim \{ \mu \} $,where $\{ \mu \}$ 
denotes the IRREP of $SU(M)$ corresponding to  the Young diagram $\{ \mu \}$.\\
 ${\it Remark}$.
If we consider the interactions of particles described by the Hamiltonian 
$H=H(N_1,..N_M)$, where $N_i$ are number operators,then
 \begin{equation}
{\cal Z}^{\Gamma}_{N}(H)=\sum_{n_{1}+ \cdots + n_{M}=N}d^{\Gamma}_{n_{1},\cdots n_{M}}
e^{-\beta H(n_1,...n_M)} .
\end{equation}
\noindent
If the algebra $\Gamma$,Eq.(2),is invariant under the permutation group 
$S_M$,then 
\begin{equation}
{\cal Z}^{\Gamma}_{N}(H)  = \sum_{\lambda;|\lambda|=N} d^{\Gamma}_{\lambda}\,
\sum_{\pi \in S_M} e^{-\beta H( \lambda_{\pi (1)},\cdots \lambda_{\pi (M)})}.
\end{equation}

{\bf Multi-level parastatistical systems}. Here we analyze multi-level para-Bose and 
para-Fermi systems since they are important examples invariant under the permutation group 
$S_M$.The operator algebra corresponding to parastatistics of order ${\it p}$ 
[1-3] is defined by trilinear relations
\begin{equation}
[a^{\dagger}_ia_j \pm a_ja^{\dagger}_i,a^{\dagger}_k]=(2/p) \delta_{jk}a^{\dagger}_i ,
\quad \forall i,j,k =1,2,...,M . 
\end{equation}
\noindent
Its Fock representation satisfies the following conditions: 
$$
<0|0>=1 \qquad  a_j|0>=0
$$
\begin{equation}
a_ia^{\dagger}_j|0>=\delta_{ij}|0> \qquad i,j=1,...M
\end{equation} 
\noindent
with $|0>$ denoting the vacuum state.
The upper (lower) sign coresponds to para-Bose (para-Fermi) algebra,and $p$ is the order 
of parastatistics.The Fock space does not contain any state with negative squared norm 
if $p$ is a positive integer [3].
The consistency condition are the following [1,3]:
\begin{equation}
[a^{\dagger}_i,[a^{\dagger}_j,a^{\dagger}_k]_{\pm}]=0 \qquad \forall i,j,k ,
\end{equation}
\noindent
where the upper (lower) sign coresponds to parabosons (parafermions). Note that Eq.(18) 
does not imply eq.(16). For $p<N$,in the N-particle space,there 
are additional null-states leading to relations not contained in Eq.(18).For 
the para-Bose and para-Fermi algebra,Eqs.(16),(17) can be presented in the form of Eq.(2), [11].\\
The matrices ${\cal A}^{p,\epsilon}_{\lambda}$, $|\lambda|=N$ can be 
calculated recursively using the following relation [11,12]:
\begin{eqnarray}
a_ia^{\dagger}_{i_1}\cdots a^{\dagger}_{i_N}|0> & = & \sum_{k=1}^{N} \delta_{ii_k} 
\epsilon ^{k-1}\, a^{\dagger}_{i_1}\cdots \hat{a}^{\dagger}_{i_k}\cdots 
 a^{\dagger}_{i_N}|0> \nonumber \\
 &-& (2/p)\sum_{k=2}^{N} \delta_{ij_k}\sum_{l=1}^{k-1}\epsilon ^{l}
a^{\dagger}_{i_1}\cdots \hat{a}^{\dagger}_{i_l}\cdots a^{\dagger}_{i_{k-1}}
a^{\dagger}_{i_l} a^{\dagger}_{i_{k+1}}\cdots a^{\dagger}_{i_N}|0> .
\end{eqnarray}
\noindent
where $\epsilon =\mp 1$ , the upper (lower) sign is for parabosons (parafermions), and 
the  $\hat{a}$ denotes omission of the corresponding operator a.

${\it Case \, M=2}$.$\,$ Two-level parastatistical systems have already been studied in Ref.[13].
But the method in [13] is not completely correct (in fact the results for 
parafermions are false -see (20) bellow).For $M=2$,from Eq.(16) it follows for parabosons 
$(a^{\dagger}_{i})^2a^{\dagger}_{j}= a^{\dagger}_{j}(a^{\dagger}_{i})^2 $,$ i,j =1,2,$
and for parafermions $(a^{\dagger}_{i})^2a^{\dagger}_{j}=2a^{\dagger}_{i}a^{\dagger}_{j}a^{\dagger}_{i}
- a^{\dagger}_{j}(a^{\dagger}_{i})^2 $,$i,j =1,2 $.\\
Then every N-particle state,$N=n_1+n_2$,$ n_1\geq n_2$, and $n_1,n_2$ fixed,can be 
expressed in terms of the following $n_2+1$ vectors [13]:
$$
(a^{\dagger}_{1})^{n_1}(a^{\dagger}_{2})^{n_2}|0> \, ;\quad
(a^{\dagger}_{1})^{n_1-1}a^{\dagger}_{2}a^{\dagger}_{1}(a^{\dagger}_{2})^{n_2-1}|0> ;\cdots
;(a^{\dagger}_{1})^{n_1-n_2}
\underbrace{a^{\dagger}_{2}a^{\dagger}_{1}\cdots a^{\dagger}_{2}a^{\dagger}_{1}}_{2n_2}|0>.
$$
\noindent
In order to find the number $d^{p,\epsilon}_{\lambda}(\leq \lambda_2 +1)$of linearly independent states  of type $\lambda$ ,we have studied the 
associate $(\lambda_2+1) \times (\lambda_2+1)$ Gram matrices up to $N=6$.For example,
for $N=6$ there are three cases, i.e.  
$\lambda_2 = 1,2,3$ with $2 \times 2$, $3 \times 3$, $4 \times 4$ types of matrices,
respectively.We found that for parabosons (with $p\geq 2$),$d_{\lambda}=\lambda_2 +1$,
i.e. the result does not depend on $p$ and coincides with the number of 
allowed IRREP's $\mu$ for fixed $\lambda_1,\lambda_2$ in the decomposition, Eq.(10).
However,for parafermions $p\geq1$ we found up to $N\leq 6$ (in contrast to [13]):
\begin{equation}
d^{p,+}_{\lambda}=\left\{ \begin{array}{lll}
0 &\mbox{if $p<\lambda_1$} \\
p-\lambda_1 +1 & \mbox{if $\lambda_1\leq p \leq \lambda_1+\lambda_2=N$} \\
\lambda_2 +1 & \mbox{if $p\geq \lambda_1+\lambda_2 =N$}
\end{array}
\right.
\end{equation}
\noindent
i.e. the result does depend on $p$ if $p<N$,in contrast to [13] and just coincides with the 
number of allowed IRREP's $\mu$ for fixed $\lambda_1,\lambda_2$ and $\lambda_1 \leq p$ 
in the decomposition Eq.(10).
Furthermore,if number of parts in $\lambda $ (the lenght of $\lambda$), 
$l(\lambda)\leq 2$, then $K_{\mu\lambda}=1$ for any of the $2\lambda_2+1$ 
allowed $\mu$, i.e. for $\mu=(\lambda_1+\lambda_2,0), \cdots ,
\mu=(\lambda_1,\lambda_2)$, [15].This means that every IRREP $\mu$ for $\lambda $, 
with $l(\lambda)\leq 2$,is filled  only with one state.Hence, 
$d^{p,\epsilon}_{\lambda}=\sum_{\mu}n^{p,\epsilon}(\mu)$ 
and sum runs over allowed IRREP's for parabosons or parafermions.\\
The  results for multiplicities  $n^{p,\epsilon}(\mu)$ of IRREP $\mu$ in $d^{p,\epsilon}_{\lambda}$ Eq.(10), 
(checked on a computer up to $N\leq 6$, $l(\lambda)\leq 2$) 
 are for parabosons $p\geq 1$:
\begin{equation}
n^{pB}(\mu)=\left\{ \begin{array}{ll}
1&\mbox{if $l(\mu)\leq min(2,p)$}\\
0 & \mbox{otherwise}
\end{array}
\right.
\end{equation}
\noindent
and for parafermions $p\geq 1$:
\begin{equation}
n^{pF}(\mu)=\left\{ \begin{array}{ll}
1&\mbox{if $l(\mu)\leq 2 \; and \; l(\mu ^T)\leq p$}\\
0 & \mbox{otherwise}
\end{array}
\right.
\end{equation}
\noindent
where $\mu ^T $ is the transposed tableau of  $\mu $. Note that the results, Eqs.(20-22), for a given 
$\lambda = (\lambda_1, \lambda_2)$ are valid 
for all $M\geq 2$.\\
Starting with Green's second quantized approach eqs.(16)-(18) and using the 
theorem proved in next subsection, we obtain the following :\\
General result:$ n^{p,\epsilon}(\mu)=1$ for all allowed IRREP's $\mu$ in the decomposition 
of $d^{p,\epsilon}_{\lambda}$,eq.(10).\\
%Hence ,we conjecture that $n(\mu)=1$ for all allowed IRREP's $\mu$ i
%in the decomposition 
%of $d_{\lambda}$ ,Eq.(10). 
In particular for  $p \geq N=|\mu|= |\lambda|$, the multiplicities and dimensions for 
parabosons and parafermions coincide,i.e. $n^{pB}(\mu)=n^{pF}(\mu)$ and 
$d_{\lambda}^{pB}=
d_{\lambda}^{pF}$.\\
For the partition function for two-level ($M=2$) parabosons of order $p \geq 2$ 
we then have:
\begin{equation}
{\cal Z}^{pB}(x_1,x_2;p)= \frac{1}{(1-x_1)(1-x_2)(1-x_1x_2)}.
\end{equation}
\noindent
i.e. it is independent of ${\it p}$ (in agreement  with [13]).
The partition function  for two-level ($M=2$) parafermions of order ${\it p}$ 
satisfies the recurrence relation,which follows from $d^{p,+}_{\lambda}$, Eq.(20):
\begin{equation}
{\cal Z}^{pF}(x_1,x_2;p)={\cal Z}^{pF}(x_1;p){\cal Z}^{pF}(x_2;p)+x_1\cdot x_2 \,
{\cal Z}^{pF}(x_1,x_2;p-2),
\end{equation}
\noindent
where 
$$
{\cal Z}^{pF}(x;p)=\frac{x^{p+1}-1}{x-1} \quad p\geq 0.
$$
\noindent
The solution of eq.(24) is given by
\begin{eqnarray}
{\cal Z}^{pF}(x_1,x_2;p) & = &\sum_{k=0}^{[\frac{p}{2}]}(x_1 x_2)^k\,{\cal Z}^{pF}(x_1;p-2k)
\, {\cal Z}^{pF}(x_2;p-2k) \nonumber \\
& = & \frac{1}{(1-x_1)(1-x_2)} \,[\, \frac{1- (x_1 x_2)^{p+2}}{1-x_1x_2} - \frac{
x_1^{p+2} - x_2^{p+2}}{x_1-x_2}\, ].
\end{eqnarray}
This result is new and differs from that in [13].In particular, we find for $p=1,2$: 
$$ {\cal Z}^{pF}(x_1,x_2;1)=(1+x_1)(1+x_2),$$
$$ {\cal Z}^{pF}(x_1,x_2;2)=(1+x_1+x_1^2)(1+x_2+x_2^2)+ x_1x_2,$$
\noindent
and for $ p\geq 3 $, ${\cal Z}^{pF}(x_1,x_2;p)$ can be easily found using Eq.(24) or Eq.(25).
In the limit $E_2 \rightarrow \infty $, $x_2\rightarrow 0 $,Eq.(23),(25)   
reduces to the partition function for one-level parabosons (parafermions).
If $p>>1$, then ${\cal Z}^{pF}(x_1,x_2;p) \approx {\cal Z}^{pB}(x_1,x_2;p)$.However,  
for  $p=\infty$ the para-Bose (para-Fermi) algebra (16) becomes Fermi (Bose) and hence
$$
{\cal Z}^{pF}(x_1,x_2;p=\infty)= \frac{1}{(1-x_1)(1-x_2)}={\cal Z}^{B}(x_1,x_2)
$$
$$
{\cal Z}^{pB}(x_1,x_2;p=\infty)=(1+x_1)(1+x_2)={\cal Z}^{F}(x_1,x_2).
$$
Generally, there is no smooth transition from generalized statistics to Bose or Fermi 
statistics,or from "higher" to "lower" statistics.\\
{\it General case $M\geq 2$.} 
Here we generalize the results of Ref.[13] to any multi-level system and the
results of Ref.[14] to parastatistics of any order following the Green's 
second quantized approach, eqs.(16)-(18).\\ 
{\it Theorem.} Let us consider the $N$-particle states of type $
1^{n_1}2^{n_2}\cdot\cdot\cdot M^{n_M}$, $\sum^{M}_{i=1} n_i = N$. If the order of Green's 
parastatistics defined by eqs.(16)-(18) is $p \geq N$, then the number $d_{\lambda}$ of linearly independent physical states is 
\begin{equation}
d_{\lambda} = \sum_{\mu}\;K_{\mu\lambda}
\end{equation}
where $K_{\mu\lambda}$ denotes the Kostka numbers and $\lambda$ is partition 
of $N$ with parts $n_1,n_2,...,n_M$.
{\it Sketch of proof}. Let us consider all possible states of the type 
$1^{n_1}2^{n_2}
\cdot\cdot\cdot M^{n_M}$, $\sum^{M}_{i=1} n_i = N$. Our aim is to write the canonical basis for this subspace which generalize the corresponding basis for $M=2$, 
Ref.[13]. \\
We shall transform any given $N$-state by using the para-Bose (para-Fermi) relations,
eq.(18) in the following form (as reduction rules in which, for simplicity, we abbreviate
$a^{\dagger}_k$ to $k$ etc.):
\begin{eqnarray}
kji &=& ikj + \epsilon \;jki - \epsilon\; ijk \nonumber\\
kki &=& -\epsilon \;ikk + (1+\epsilon) \;kik, \;\;\;\;\; 1\leq i,j < k 
\end{eqnarray}
\noindent
(here ${\epsilon} =+/-$ refers to parafermions/parabosons).
We first start shifting the rightmost index $M$ to the right. This will end with $M$ either in the last or next to last right position $\ast \,M$ or 
$\ast \,Mi$, $i < M$. Then we apply the same procedure to the second 
rightmost index $M$ by shifting it to the right as before. 
After moving all of $M$'s, we end up with states like $\ast\,M^{n_M}$, 
$\ast\,M\;i_1\;M^{n_{M}-1}, \cdot\cdot\cdot , \ast\,M\;i_1\;M\;i_2\cdot\cdot\cdot M\;i_{n_M}$, i.e. 
with $n_{M}+1$ types of states.\\
By applying the approach of Bergman, Ref.[16] to the relations (27), we find  
new important relations (no further relations except these exist if $p \geq N$)
\begin{equation}
kjki = kikj + jkik - ikjk, \;\;\;\; k > j \geq i,
\end{equation}
\noindent
which together with (27) ensure the diamond condition for the para-Fermi/para-
Bose algebras.
By applying (28) to the above types of states we further reduce them to the 
states satisfying $1\leq i_1 \leq i_2 \leq \cdot\cdot\cdot \leq i_{n_M} < M$.\\
After exhausting $M$'s we perform the same procedure on $(M-1)$'s , $(M-2)$'s
,.. and finally on $2$'s. In this way we obtain a canonical basis which 
generalizes the basis for two-level system of the previous subsection. We point out that the 
dimensions $d_{\lambda}$ satisfy the same recursion relation as coefficients $a_{\lambda}$, (
$a_{\lambda} = \sum_{\mu}\,K_{\mu\,\lambda}$) in the expresion $\prod_{i=1}^{M}1/(1-x_i)\;\prod_{i < j}^{M} 1/
(1-x_i x_j) = \sum_{\lambda}\,a_{\lambda}\, m_{\lambda}$. 
Hence, we conclude that the eq.(26) in the theorem is proved. (Alternatively, 
one
obtains one to one correspondence between the states in canonical basis and 
terms in the expansion of $\sum_{\mu} s_{\mu}(x_1,...,x_M)$.\,)\\
Comparing the eq.(26) with eq.(10), we obtain $n(\mu)=1$ for all allowed IRREP's
 $\mu$, i.e. $l(\mu) \leq M$ for para-Bose and para-Fermi case.\\
Furthermore, if $p < N$ then the Green's parastatistics imply that only IRREP's with $l(\mu) \leq p$ for para-Bose and $l(\mu^{T}) \leq p$ for para-Fermi are
allowed. These are additional restrictions on the multiplicities $n(\mu)$ in 
eq.(10). We point out that these restrictions are not put by hand, but they are
intrinsicaly contained in eq.(16). The complete proof of the theorem and 
corrolaries will be presented elsewhere.\\
This theorem enables a simple generalization of Eq.(23).The partition
 function for multi-level parabosons for $ p\geq M $ is:
\begin{equation} 
{\cal Z}^{pB}(x_1,...x_M;p\geq M)=\sum_{\lambda}
s_{\lambda}(x_1,...x_M)=\prod_{i=1}^{M}\frac{1}{(1-x_i)}
\prod_{i<j}^{M}\frac{1}{(1-x_ix_j)}.
\end{equation}
\noindent
The  identity used in Eq.(29) can be found in [15].The results  for parastatistical systems of 
order $p=2$ [14] can be easily obtained from Eqs.(10,11) and our eq.(26).
We point out that the parastatistics in the first quantized approach of Ref.[17], [18] leads to the same partition functions as we have found. However, these two parastatistics basicaly differ:
(i) since within the first quantized approach [17],[18], the Green's trilinear 
commutation
 relations , eq.(18) cannot be obtained, and
(ii) the probabilites of finding IRREP $\Gamma_c$ in $\Gamma_a \otimes \Gamma_b
 = \sum_c \Gamma_c$ and finding $\Gamma_a \otimes \Gamma_b$ in $\Gamma_c = 
\sum_{(a,b)} \Gamma_a \otimes \Gamma_b$ calculated within the approach of 
[17],[18] would be completely different from those 
calculated in the second quantized approach. 
Hence these two approaches are not equivalent. The complete analysis will be 
published separately.\\
${\bf Anyonic-like \, multi-level \, systems}$. Here we study another class of multi-level systems 
described by algebras with Fock representations,which possess anyonic-like statistics and 
are not invariant under the permutation group. We do not assume any symmetry requirements 
. Instead, we assume for $i \neq j$:
\begin{equation} 
a_ia^{\dagger}_j =e^{i\phi_{ij}}a^{\dagger}_ja_i ,\qquad \phi_{ij}=- \phi_{ji} \in {\bf R},
\qquad i\neq j ,\quad i,j=1,2...M
\end{equation}
%$$
%\phi_{ij}=- \phi_{ji} \in {\bf R}.
%$$
%$$
%i\neq j ,\qquad i,j=1,2...M
%$$
and  $a_i$, $a^{\dagger}_i$ are adjoint to each other.It is easy to show that 
$d_{n_{1},\cdots n_{M}}=1$ for any set of mutually different indices $i_{1},\cdots i_{N}$. In 
this case, any monomial state consisting of permuted indices $i_{1},\cdots i_{N}$ is ,up to 
unit phase, equal to ($a^{\dagger}_{i_{1}}\cdots a^{\dagger}_{i_{N}}|0>$). Therefore, we find 
the anyonic-like commutation relations [5]:
\begin{equation} 
a_i^{\dagger}a^{\dagger}_j = e^{-i\phi_{ij}}a^{\dagger}_ja_i^{\dagger},
\qquad a_ia_j = e^{-i\phi_{ij}}a_j a_i \qquad i\neq j.
\end{equation}
We have assumed no specific relation $a_ia^{\dagger}_i=\Gamma (a^{\dagger},a)$ for the same indices. 
The independent number operators $N_i$ are assumed. Then for a multi-level anyonic system 
with the free Hamiltonian $H_0=\sum_{i=1}^{M} E_i \, N_i$, the partition function is 
\begin{equation} 
{\cal Z}(x_1,...x_M;\phi_{ij})=\prod_{i=1}^{M}{\cal Z}(x_i,p_i),
\end{equation}
\noindent
where ${\cal Z}(x;p)=(1-x^{p+1})/(1-x)$ is the one-level partition function of a single 
(parafermion-like) oscillator satisfying $a^{p+1}=0$, $p\in {\bf N}$. Such a system is physically 
equivalent to the set of M commuting ( or anticommuting ) single-mode 
generalized oscillators $\tilde{a}_i^{\dagger}$, $\tilde{a}_i$. The two sets of operators $\{ a_i \}$ 
and $\{ \tilde{a}_i \}$ are connected by a generalized Jordan-Wigner transformation [19].
\\ Starting from  Haldane's definition of generalized exclusion statistics [20], 
Karabali and Nair [21], 
using additional assumptions derived an operator algebra of the type given by eq.(30) with 
$\phi_{ij}=\frac{\pi}{p+1} sign (i-j)$, $i\neq j$ and  $a_i^{p+1}=0 $, for  $\forall i$. From our results 
it follows that the partition function corresponding to the system described by the Karabali-Nair 
algebra with the free Hamiltonian $H_0=\sum_{i=1}^{M} E_i \, N_i $ is 
\begin{equation} 
{\cal Z}(x_1,...x_M;p)=\prod_{i=1}^{M}\frac{1-x_i^{p+1}}{1-x_i}.
\end{equation}
The system with this partition function was analyzed in [22]. However it does not reproduce the 
Haldane-Wu interpolating distribution.\\
Finally,let us mention algebras with no independent number operators $N_i$, but with a well-defined 
total number operator N , $[N,a_{i}^{\dagger}]=a^{\dagger}_{i}$,
$[N,a_{i}]=-a_{i}$  for $i,j=1,2,..M$. Let us consider a free Hamiltonian of the form 
$H_0=E\, N$. Then the partition function is 
\begin{equation} 
{\cal Z}^{\Gamma}(x_1,...x_M)=\sum_{N=0}^{\infty} D^{\Gamma}(M,N)e^{-\beta E N},
\end{equation}
\noindent
where $D^{\Gamma}(M,N)$ is number of independent states 
($a^{\dagger}_{i_{1}}\cdots a^{\dagger}_{i_{N}}|0>$), \\ $i_{1},\cdots, i_{N}=1,...,M$. If 
$a_ia_j = a_j a_i$, then $ 0\leq D(M,N) \leq D^B(M,N)$. 
Such an example is the system derived from the Calogero-Sutherland model 
analyzed in [23].
A complete analysis including thermodynamic properties of this and other generalized 
multi-level systems will be treated separately.

%%%%%%%%%%%%%%%%%%%%%%%%%%%%%%%%%%%%%%%%%%%%%%%%%%%%%%%%%%%%%%%%%%%%%%%%%%%%%%%
%%%%%%%%%%%%%%%%%%%%%%%%%%%ACKNOWLEDGEMENTS%%%%%%%%%%%%%%%%%%%%%%%%%%%%%%%%%%%%

${\bf Acknowledgements}$. The authors thank 
I. Dadi\c $\,$ and  M. Milekovi\c $\,$ for useful discussions.

%%%%%%%%%%%%%%%%%%%%%%%%%%%%%%%%%%%%%%%%%%%%%%%%%%%%%%%%%%%%%%%%%%%%%%%%%%%%%%
%%%%%%%%%%%%%%%%%%%%%%%%%%%REFERENCES%%%%%%%%%%%%%%%%%%%%%%%%%%%%%%%%%%%%%%%%%
\newpage
\baselineskip=24pt
{\bf REFERENCES}
\begin{description}

\item{[1]}
H.S.Green, Phys.Rev. 90 (1953) 170.
\item{[2]}
O.W.Greenberg and A.M.L.Messiah, Phys.Rev. 138 B (1965) 1155; 
J.Math.Phys. 6 (1965) 500.
\item{[3]}
Y.Ohnuki and S.Kamefuchi, Quantum field theory and parastatistics 
( University of Tokio Press, Tokio, Springer, Berlin, 1982).
\item{[4]}
O.W.Greenberg, Phys.Rev.D 43 (1991) 4111; R.N.Mohapatra, Phys.Lett. B242 (1990) 407.
\item{[5]} 
S.Meljanac and A.Perica, Mod.Phys.Lett.
 A 9 (1994) 3293;  J. Phys.A :Math.Gen. 27 (1994) 4737;
	V.Bardek, S.Meljanac and A.Perica, Phys.Lett. B 338 (1994) 20.
\item{[6]}
A.B.Govorkov, Nucl.Phys. B 365 (1991) 381.
\item{[7]}
J.M.Leinaas and J.Myrheim,  Nuovo Cim. 37 (1977) 1; F.Wilczek, Phys.Rev.Lett. 48 (1982) 1144.
\item{[8]}
V.Bardek, M.Dore\s i\c $\,$ and S.Meljanac, Phys. Rev. D 49 (1994) 3059 ; 
V.Bardek, M.Dore\s i\c $\,$ and S.Meljanac, Int.J.Mod.Phys. A 9 (1994) 4185;
A.K.Mishra and G.Rajasekaran, Mod.Phys.Lett. A 9 (1994) 419.
\item{[9]}
S.B.Isakov, Int.J.Theor.Phys. 32 (1993) 737.
\item{[10]}
S.Meljanac , M.Milekovi\c $\,$ and S.Pallua, Phys.Lett. 328B (1994) 55.
\item{[11]}
S.Meljanac and M.Milekovi\c , Unified view of multimode algebras with Fock-like 
representations, Int.J.Mod.Phys. A 11 (1996) 1391.
\item{[12]}
A.B.Govorkov, Theor.Math.Phys. 98 (1994) 107.
\item{[13]}
A.Bhattacharyya,F.Mansouri,C.Vaz and L.C.R.Wijewardhana, 
Phys.Lett. 224B (1989) 384.
\item{[14]}
P.Suranyi, Phys.Rev.Lett.  65 (1990) 2329 .
\item{[15]}
I.G.Macdonald, {\it Symmetric functions and Hall polynomials} (Claredon, 
Oxford, 1979).
\item{[16]}
G.M.Bergman, Adv.Math. 29 (1978) 178.
\item{[17]}
S.Chaturvedi, Canonical partition functions for parastatistical system 
of any order, preprint hep-th/9509150.
\item{[18]}
 A.P.Polychronakos, Path integrals and parastatistics, preprint hep-th/9603179.
\item{[19]}
M.Dore\s i\c ,S.Meljanac and M.Milekovi\c, Fizika 3 (1994) 57.
\item{[20]}
F.D.M.Haldane, Phys.Rev.Lett.  67 (1991) 937 ; Y.S.Wu, Phys.Rev.Lett.  73 (1994) 922.
\item{[21]}
D.Karabali and V.P.Nair, Nucl.Phys.B  438 [FS] (1995) 551.
\item{[22]}
S.I.Ben-Abraham, Am.J.Phys. 38 (1970) 1335.
\item{[23]}
 A.P.Polychronakos,  Phys. Rev. Lett. 69  (1991) 703; 
 L.Brink , T.H. Hanson, S.Konstein and M.A.Vasiliev,  Nucl.Phys. 401B (1993) 591.
\end{description}
\end{document}